\documentclass[pre,twocolumn,floatfix, superscriptaddress,a4paper,nofootinbib]{revtex4-1}
\usepackage[utf8]{inputenc}
\usepackage{graphics}
\usepackage{amsmath,bm,graphicx}
\usepackage{amssymb}
\usepackage{amsfonts}
\usepackage{graphicx}
\usepackage{epstopdf}
\usepackage{amssymb}
\usepackage{mathrsfs}
\usepackage{amsmath}
\usepackage{color}
\usepackage{latexsym}
\usepackage{amsfonts}
\usepackage{hyperref}
\usepackage{subfigure}
\usepackage{wasysym}
\usepackage{dcolumn}
\usepackage{bm}
\usepackage{natbib}
\usepackage{cancel}

\begin{document}

\title{Charge polarization, local electroneutrality breakdown and eddy formation due to electroosmosis in varying-section channels}

\author{Mauro Chinappi}
\email{mauro.chinappi@uniroma2.it}
\affiliation{Dipartimento di Ingegneria Industriale, Universit\`a di Roma Tor Vergata, via del Politecnico 1, 00133 Roma}
\author{Paolo Malgaretti}
\email[Corresponding Author : ]{malgaretti@is.mpg.de }
\affiliation{Max-Planck-Institut f\"{u}r Intelligente Systeme, Heisenbergstr. 3, D-70569
Stuttgart, Germany}
\affiliation{IV. Institut f\"ur Theoretische Physik, Universit\"{a}t Stuttgart,
Pfaffenwaldring 57, D-70569 Stuttgart, Germany}

\begin{abstract}
We characterize the dynamics of an electrolyte embedded in a varying-section channel under the action of a constant external electrostatic field. By means of molecular dynamics simulations we determine the stationary density, charge and velocity profiles of the electrolyte. Our results show that when the Debye length is comparable to the width of the channel bottlenecks a concentration polarization along with two eddies sets inside the channel. Interestingly, upon increasing the external field, local electroneutrality breaks down and charge polarization sets leading to the onset of net dipolar field. This novel scenario, that cannot be captured by the standard approaches based on local electroneutrality, opens the route for the realization of novel micro and nano-fluidic devices.
\end{abstract}
\maketitle
The transport of ions, molecules and polymers across constrictions such as  pores, membranes or varying-section micro- nano-channels is crucial for several biological as well as synthetic systems. For example, in biological cells ion channels control the uptake of ions from the environment~\cite{hille2001ion} whereas, in resistive pulse sensing techniques, the interactions of colloidal particles~\cite{Willmott2015} or macromolecules~\cite{bayley2000stochastic,celaya2017label,bonome2015multistep,asandei2017nanoscale,chinappi2018protein} with the nano- or micro-pore are measured from the variation of the electric conductance of the pore induced by the presence of the particle. 
Moreover, electro-osmotic flux can play a crucial role in molecule capturing in nanopores~\cite{asandei2016electroosmotic,boukhet2016probing,huang2017electro}.
Recent studies have shown that ionic transport and electro-osmotic flow in micro- and nano-fluidic circuitry can be controlled by tuning the geometry of the micro- nano-channel\cite{malgaretti2015,malgaretti2016,Bolet2018}. Indeed, conical~\cite{Guo2013,Balme2017,experton2017ion} or heterogeneously charged~\cite{Bocquet2013} pores have been used to realize nano-fluidic diodes and to rectify electro-osmotic flows~\cite{Keyser2015,Bacchin2018}. Moreover, periodic varying-section channels have been used to realize nanofluidic transistors~\cite{Siwy2008}.
Similarly, recent contributions have shown that the shape of the confining vessel and the boundary conditions therein imposed can be exploited to control the flow. Indeed, electro-osmotic transport can be strongly enhanced by grafting charged brushes on the channel walls~\cite{Chen2017} and by hydrophobic surfaces~\cite{Vinogradova2015}.

\begin{figure}[t!]
\centering
\includegraphics[width=0.49\textwidth]{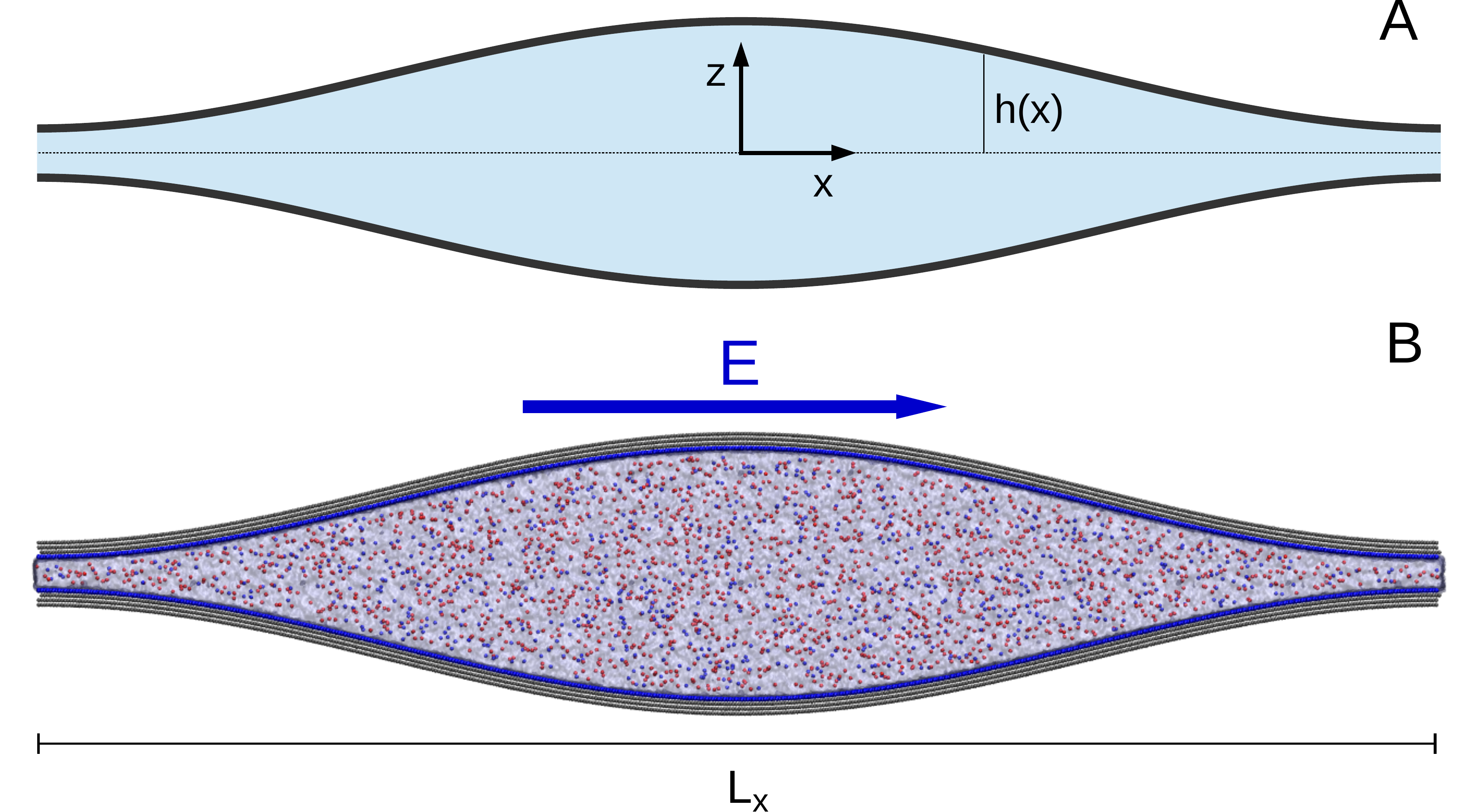}
\caption{\label{fig:sis}
System set-up. A) Sketch of the system. The electrolyte solution (light blue) 
is confined between two curved charged walls. Half section is
$h(x) = h_0 + h_1 \cos{\frac{2 \pi x}{L_x}}$.
Periodic boundary conditions are applied in the x and y direction. 
B) Snapshot of molecular dynamics set-up.
The solid walls (gray) are constituted by Lennard-Jones atoms. 
The atoms of the layer exposed to the liquid are charged (blue).
Water molecules are not reported while blue and red spheres represent
$K^+$ and $Cl^-$ ions.
An external electrical field parallel to the x-axes in applied.
}
\end{figure}

\begin{figure*}[h]
 \includegraphics[width=0.49\textwidth]{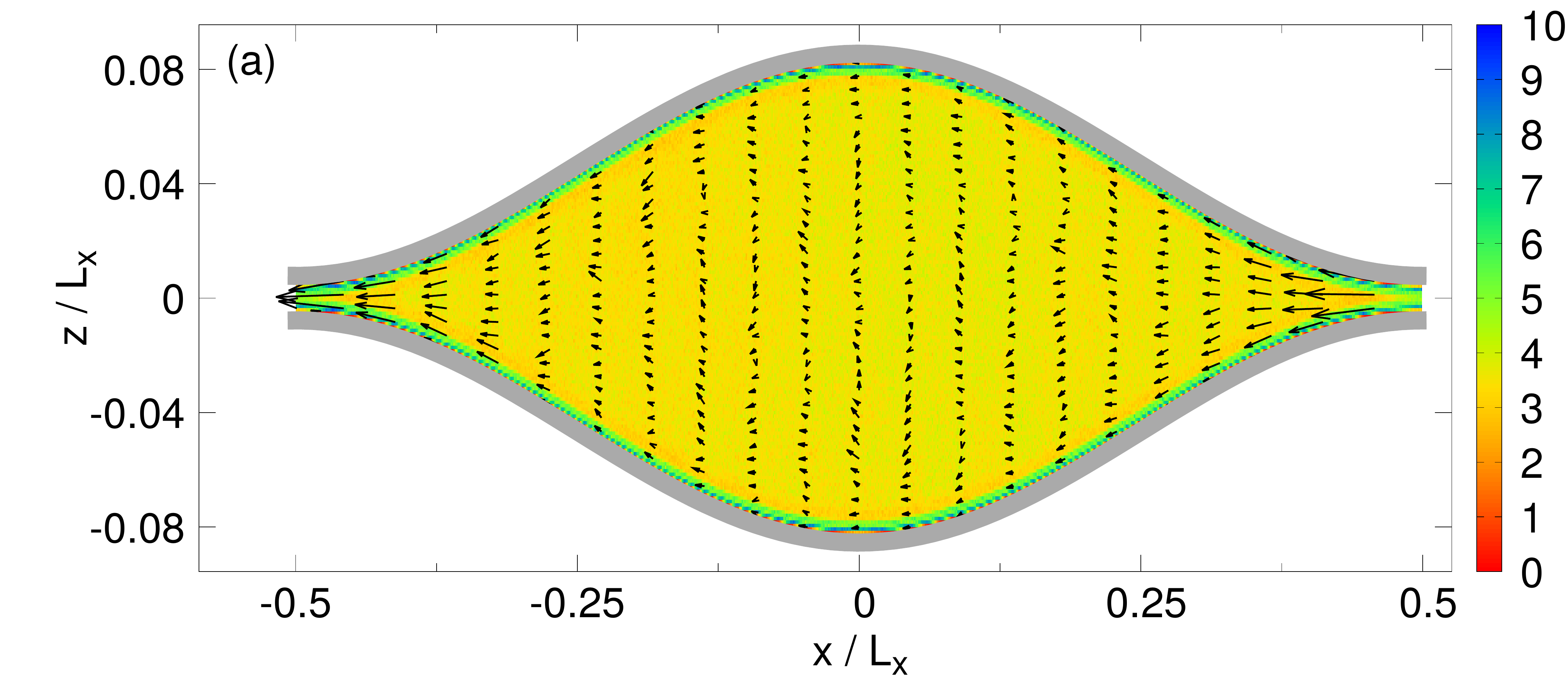} 
 \includegraphics[width=0.49\textwidth]{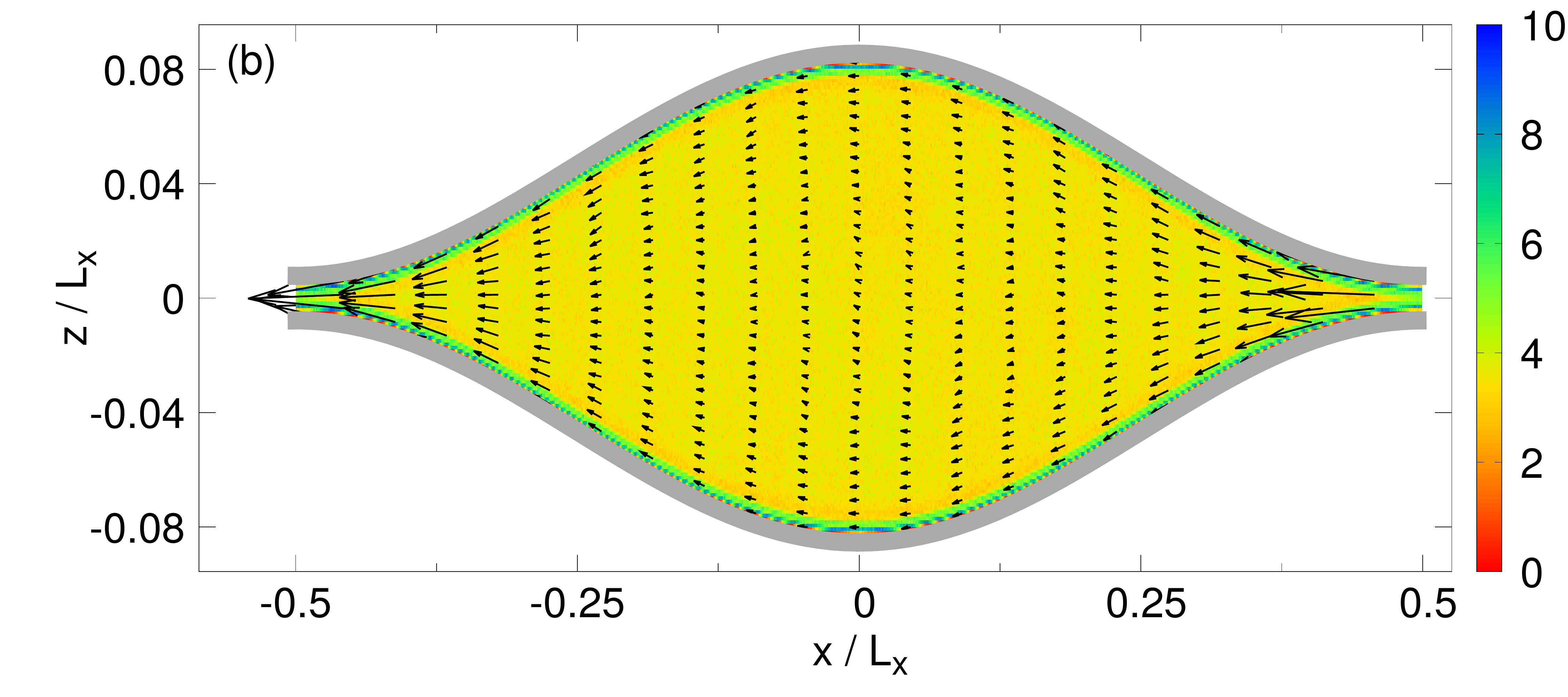}\\
 \includegraphics[width=0.49\textwidth]{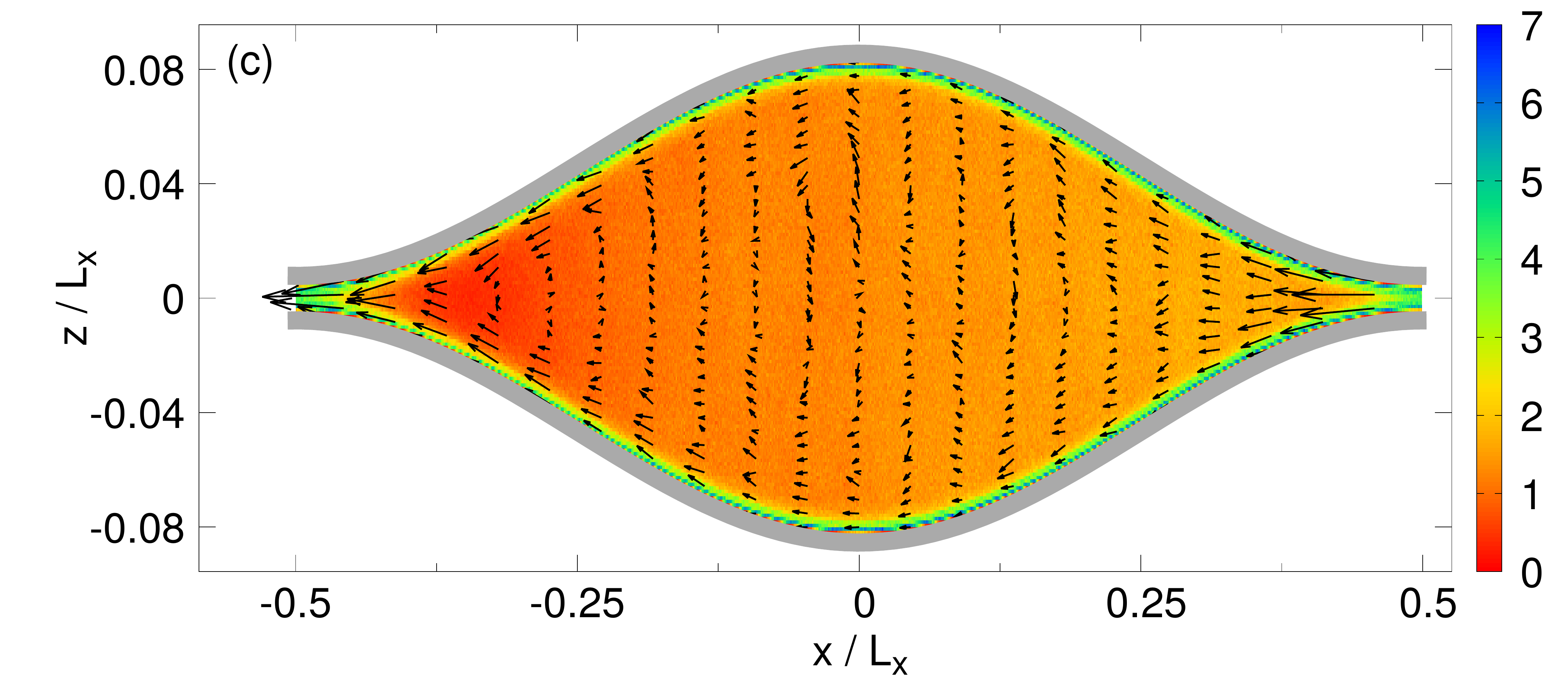}
 \includegraphics[width=0.49\textwidth]{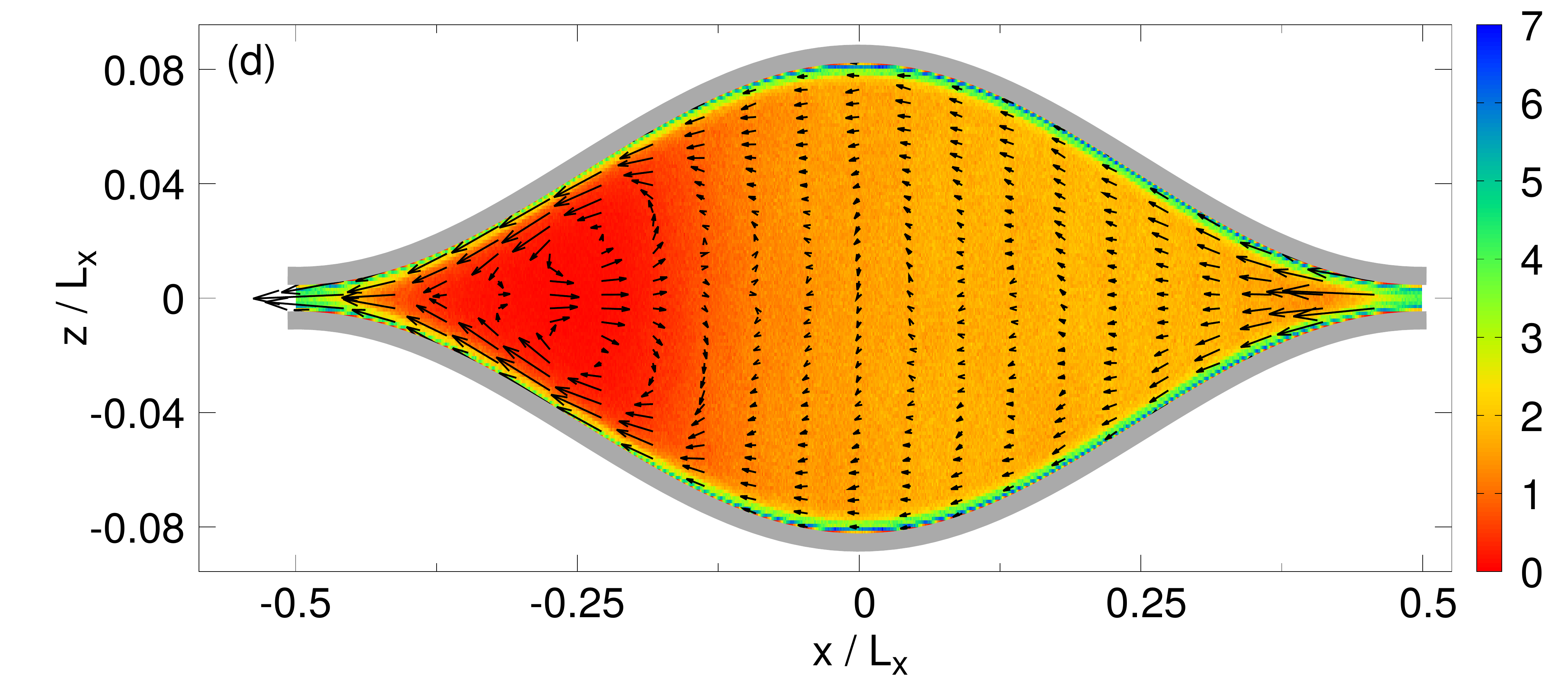} \\
 \includegraphics[width=0.49\textwidth]{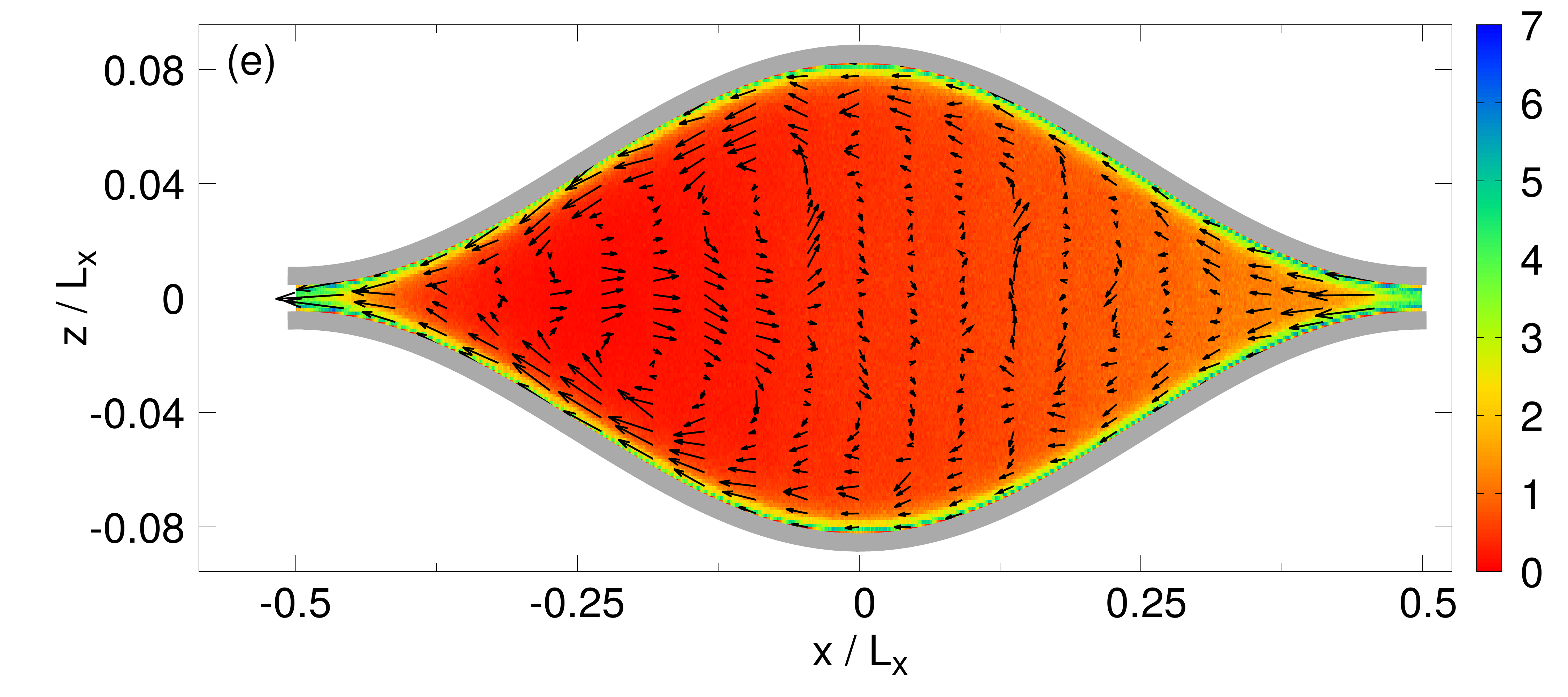}
 \includegraphics[width=0.49\textwidth]{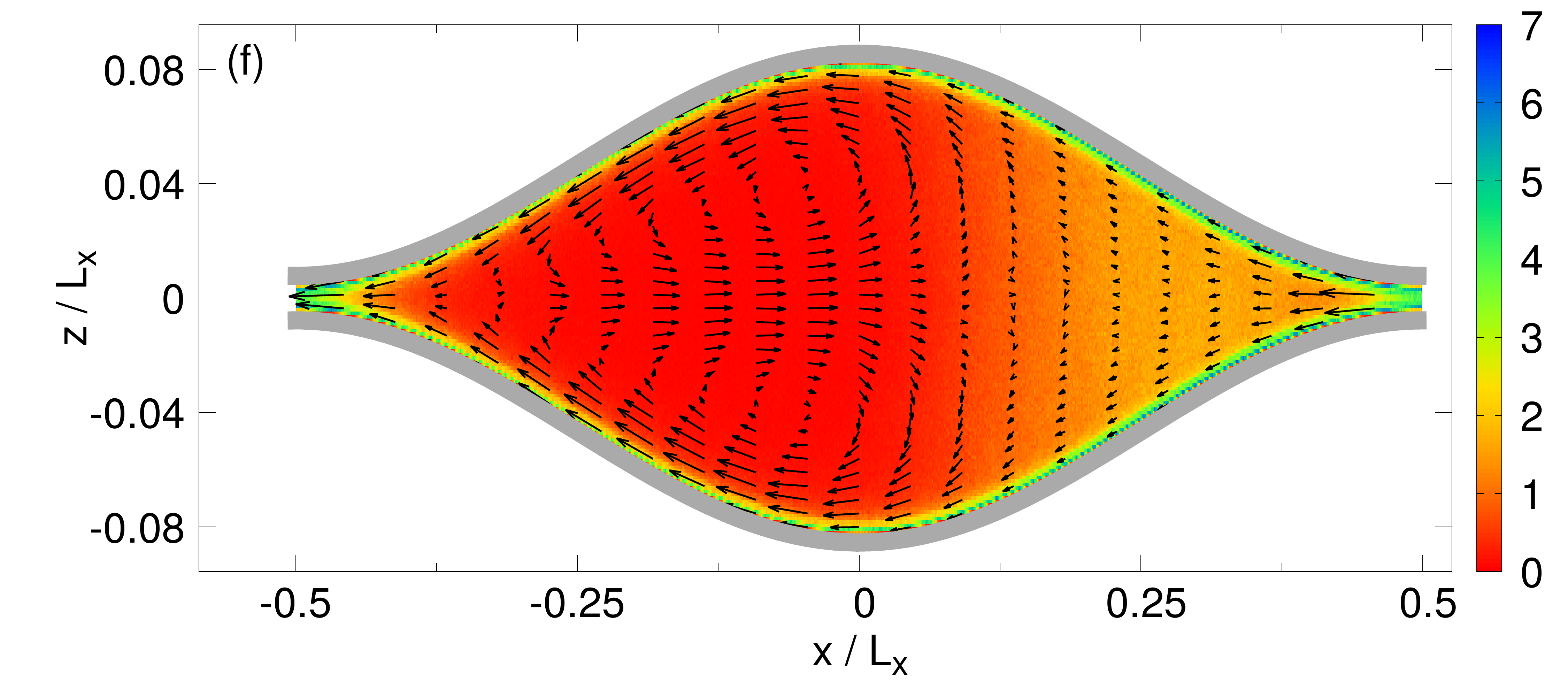}
 \caption{Net concentration of $Cl^-$ and velocity profile (arrows) 
          for external fields corresponding to potential drop $\Delta V =0.6$ (left) and $\Delta V =2.4$~V, 
          (right). 
          Panels a), b) (first row) refer to ion concentration $\rho_0 =3$M, 
          panels c), d) (central row) refer to $\rho_0 = 1$M and 
          panels e), f) (bottom row) refer to $\rho_0=0.3$M. 
          For clarity sake, the arrows of the velocity profile 
          are rescaled in each panel.
          The upper and bottom Grey solid lines indicates the channel walls.}
 \label{fig:vel-prof}
\end{figure*}
In this contribution, we study, via molecular dynamics simulations the electro-osmotic flow of an electrolyte embedded in a varying-section channel.
The advantage of our approach, as compared to others based on the solution of some continuum models~\cite{Rubinstein2000,Park2006,Mani2009,Mani2011,Yossifon2014}, is that the ionic densities are left free to relax according to the interactions among ions and between ions and the channel walls. Therefore our approach allows us to critically discuss, for example, the local electroneutrality assumption and its possible breakdown.
Our results show that when the systems is driven by a constant electrostatic field acting along the longitudinal axis of the channel, inhomogeneous ionic and charge densities are induced due to the  variations of local channel section. 
This phenomenon is similar to concentration polarization (CP) reported for electrolytes transported across ionic-selective membranes~\cite{Rubinstein2000,Mani2009,Mani2009-2} (see also Ref.~\cite{Demekhin2012,Pourcelly2014} for recent reviews) i.e., in \textit{open circuit} conditions and under severe modulations of channel section~\cite{Mani2009,Mani2009-2,Yossifon2014}. Interestingly, our results show that CP can be obtained also for \textit{close circuit} conditions and for smooth variations of channel section. In particular, we observe CP when the Debye length is comparable with the channel bottlenecks, i.e., in the entropic electrokinetic regime~\cite{Malgaretti2014}. 
Concerning local electroneutrality, our numerical results confirm that even in the entropic electrokinetic regime, the local electroneutrality \textit{assumed} by previous works~\cite{Mani2011,Pourcelly2014,Yossifon2014,Leese2014} is fulfilled for mild values of the external field. 
However, upon increasing the external field our numerical simulations show that local electroneutrality breaks down and charge polarization (QP) sets, leading to the onset of a net dipolar contribution to the electrostatic field. Interestingly, a similar phenomenon has been recently observed for pressure-driven flows across conical pores~\cite{Jubin201721987}.
\paragraph*{Results}
We study the electro-osmotic flow (see Suppl. Mat. for the details) of a KCl water solution across a channel with half section
\begin{equation}
 h(x)=h_0+h_1\cos(2\pi x/L_x)
 \label{eq:channel}
\end{equation}
with average section $h_0=38.5$\AA~ and modulation $h_1=32.5$\AA~ and periodicity 
$L_x=836.6$\AA, see Fig.~\ref{fig:sis}. Channel section is constant along the $y$ direction, with thickness $L_y=60.37$\AA, and periodic boundary conditions are applied along $x$ and $y$. Both channel walls are covered with a constant charge density $\sigma=0.15\frac{C}{m^2}$.
The system is globally electroneutral since we compensated the wall
charge with additional $Cl^-$ ions.
After equilibration, a homogeneous and constant 
external electric field ${\bf E} = (E_x,0,0)$ is applied to the whole system.

We begin our analysis by focusing on the case of larger ionic concentration $\rho_0 \simeq 3 M$. 
The density profile of both $Cl^-$ and $K^+$ is expected to decay over a length scale comparable to the Debye length, 
$\lambda=\sqrt{\frac{\varepsilon}{\beta (ze)^2 \rho_0}}$, where $e$ is the elementary charge, $z$ is the valence 
of the ions ($z=1$, in our case), $\beta= ( k_B T )^{-1}$ with $k_B$ the Boltzmann constant and $T$ the absolute temperature and $\varepsilon = 80\cdot \varepsilon_0$ is the dielectric constant with 
$\varepsilon_0$ the vacuum dielectric constant.
For $3M$ ionic solution we estimate a Debye length of $\lambda\simeq 1.8$\AA~ 
and therefore 
$\lambda \ll h_{\text{min}}$
being 
$h_{\text{min}}=h_0-h_1=6$\AA~ the half-section calculated at channel bottlenecks. In such a case,
since there is only a small overlap between the Debye layers of the two facing walls, 
we do not expect the onset of any entropic electrokinetic effects~\cite{Malgaretti2014}. 
As expected, the accumulation of $Cl^-$ ions at the positively charged solid wall
induces and electro-osmotic flow, opposed to the direction of the external electrostatic field, that is almost symmetric with respect to the $z-$axis (see panels a)-b) of Fig.~\ref{fig:vel-prof}). 
\begin{figure}
 \centering
 \includegraphics[width=0.4\textwidth]{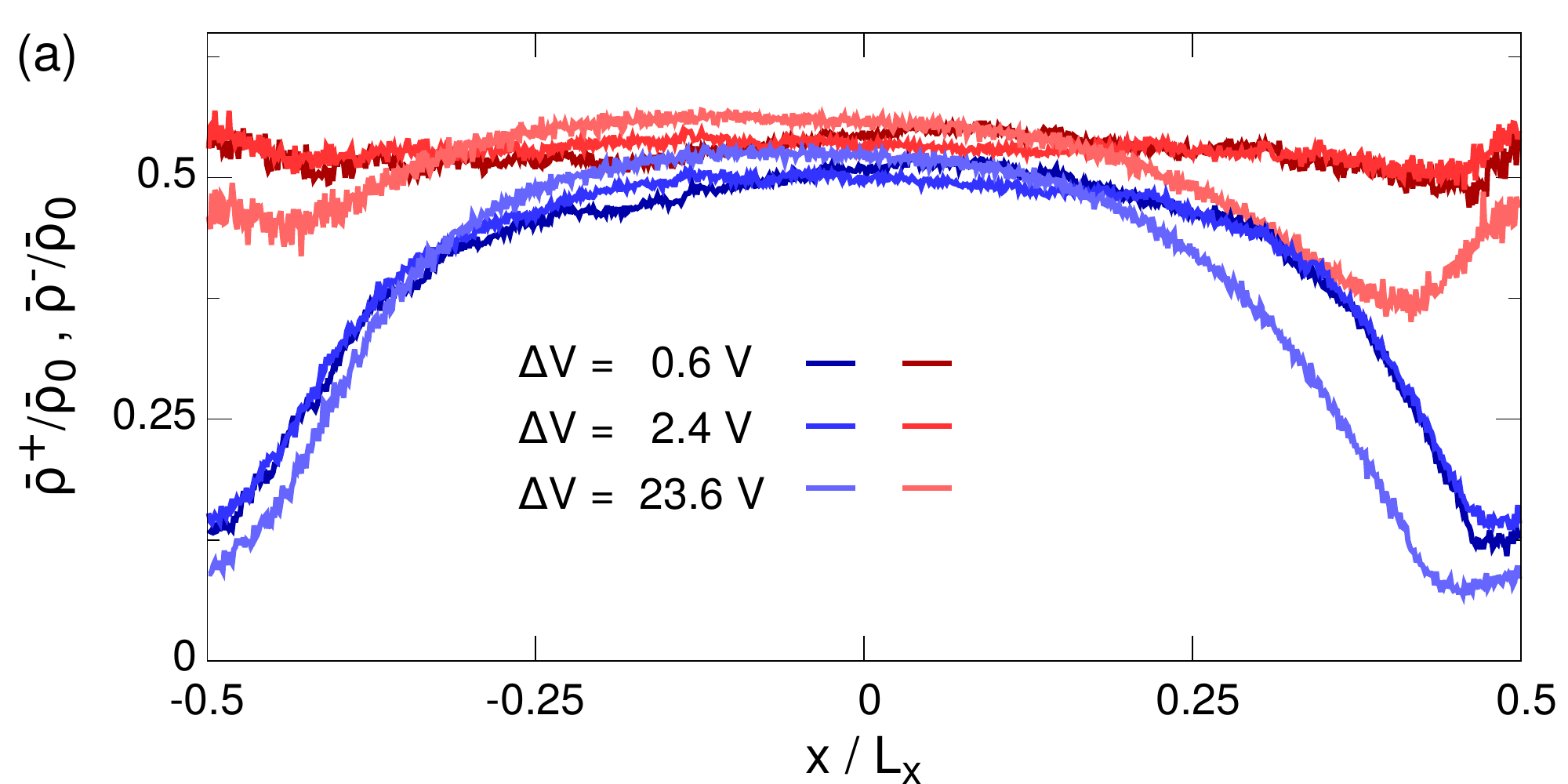}
 \includegraphics[width=0.4\textwidth]{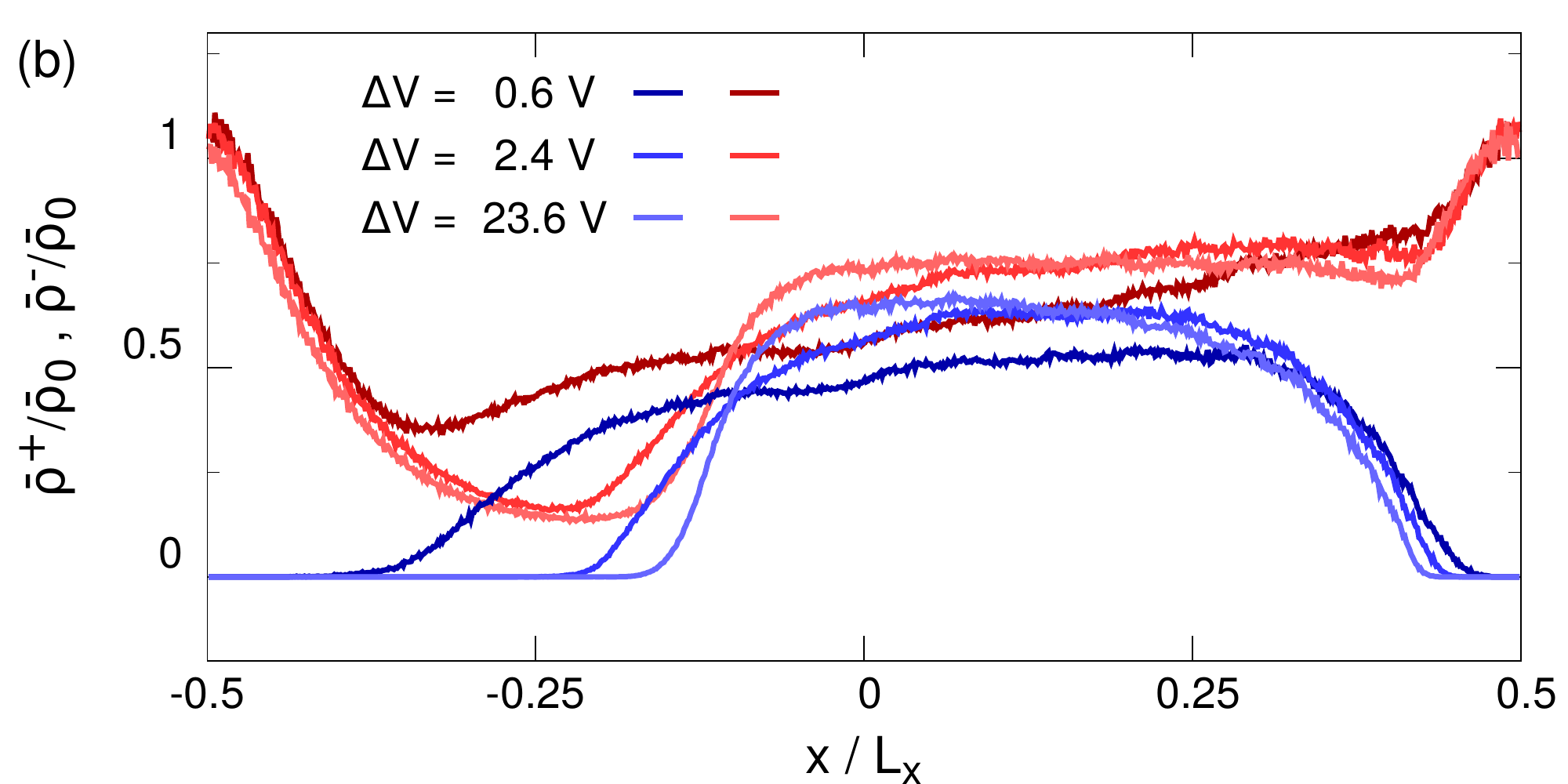}
 \includegraphics[width=0.4\textwidth]{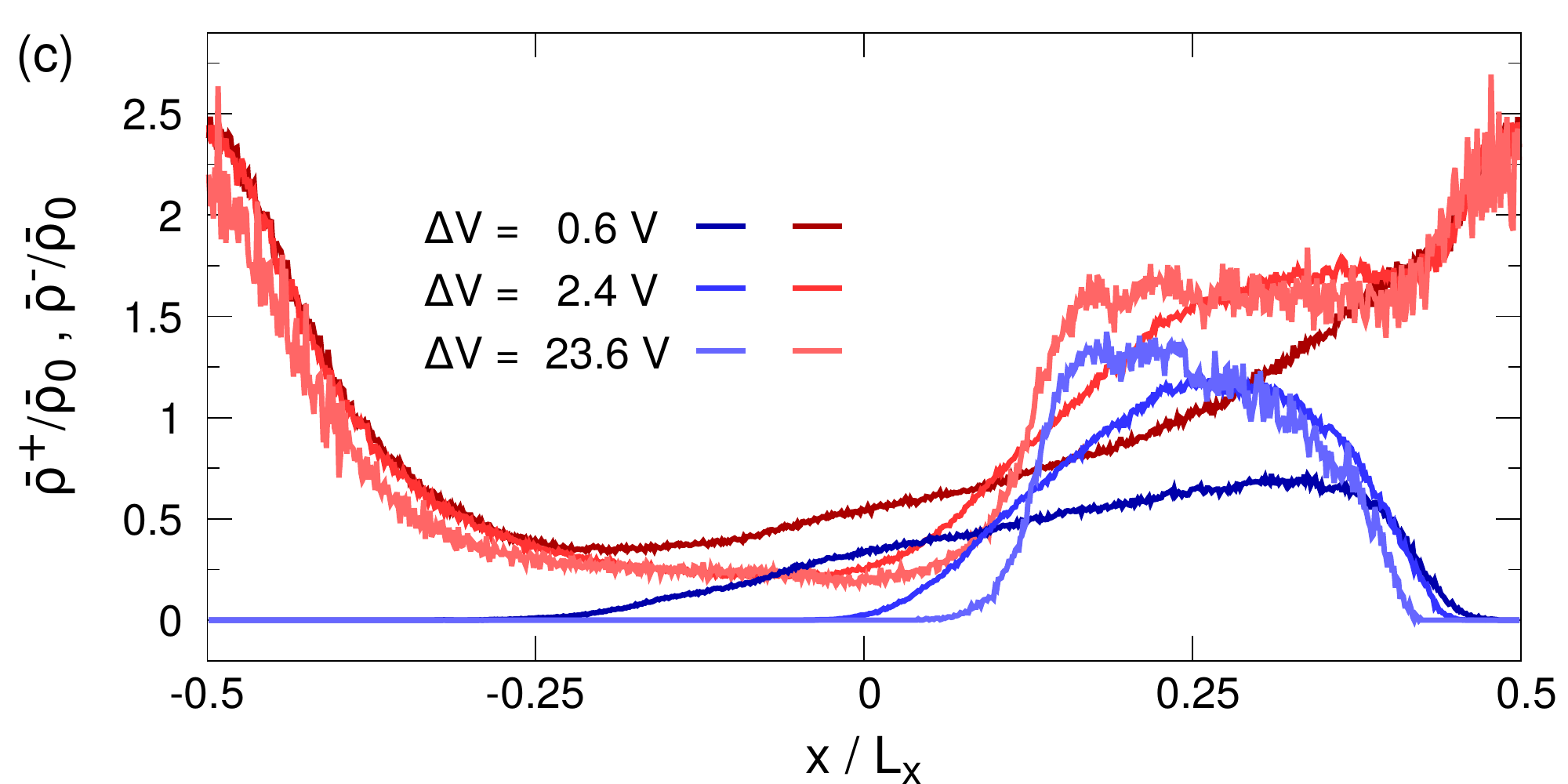}
 \caption{$\bar \rho^+(x)$ (blue) and $\bar \rho^-(x)$ (red) 
 curves as a function of the longitudinal position 
for $\rho=3M$ (panel a), $\rho=1 M$ (panel b) and $\rho=0.3 M$ (panel c). 
$\bar \rho^+(x)$ and $\bar \rho^-(x)$ are calculated via equation 
(\ref{eq:barrho}).
}
 \label{fig:profxs}
\end{figure}

We have then increased the Debye length by reducing the ionic concentration $\rho_0$. 
For $\rho_0 \sim 1M$ we have $\lambda\simeq 3.1$\AA. 
Hence, the Debye layers of
the facing walls overlaps in the narrower sections 
($\lambda/h_{\text{min}} \sim 1$, $h_{\text{min}}=6$\AA) highlighting that the system is within the entropic electrokinetic regime~\cite{Malgaretti2014}. 
Panel b) of Fig.~\ref{fig:vel-prof} shows that, for $\Delta V = 2.4$~[V], the ionic densities are quite affected by the flow. In particular, 
$Cl^-$ concentration largely increases in the channel bottleneck, see 
e.g. the region $x/L_x \in (0.45,0.5)$ in Fig.~\ref{fig:vel-prof}b. 
This increase in $Cl^-$ concentration is associated to $K^+$ depletion,  
that will be discussed more in details in next paragraphs.
This feature is associated to the onset of eddies in the electro-osmotic velocity profile, as shown in panels c)-d) of Fig.~\ref{fig:vel-prof}. These eddies form for sufficiently large driving forces. Indeed, while for $\Delta V = 0.6$~[V] we do not observe major discrepancies with the previous case, for  $\Delta V = 2.4$~[V] two eddies form inside the channel.
Interestingly, such eddies break down the left-right symmetry of the channel, 
for instance, Fig.~\ref{fig:vel-prof}d, shows that the eddies 
are shifted in the direction of the volumetric fluid flow, i.e. negative
$x$ in our reference frame. 
This occurrence is in contrast to the prediction obtained in linear regime~\cite{Malgaretti2014} 
for which the eddies center is in the channel center, $x/L_x=0$.
Then we further increase the Debye length by setting  $\rho_0 \sim 0.3M$, for which $\lambda\simeq 5.6$\AA~. 
In such a regime panels e)-f) of Fig.\ref{fig:vel-prof} show that the zone in which $K^+$ are depleted is enhanced as compared to the previous cases. Moreover, comparing panels e)-f) to c)-d) in Fig.\ref{fig:vel-prof} we notice that, for $\rho_0 \sim 0.3M$, the onset of the eddies occurs for smaller values of the external force as compared to $\rho_0 \sim 1M$. 

%
%

\begin{figure}
 \centering
 \includegraphics[width=0.4\textwidth]{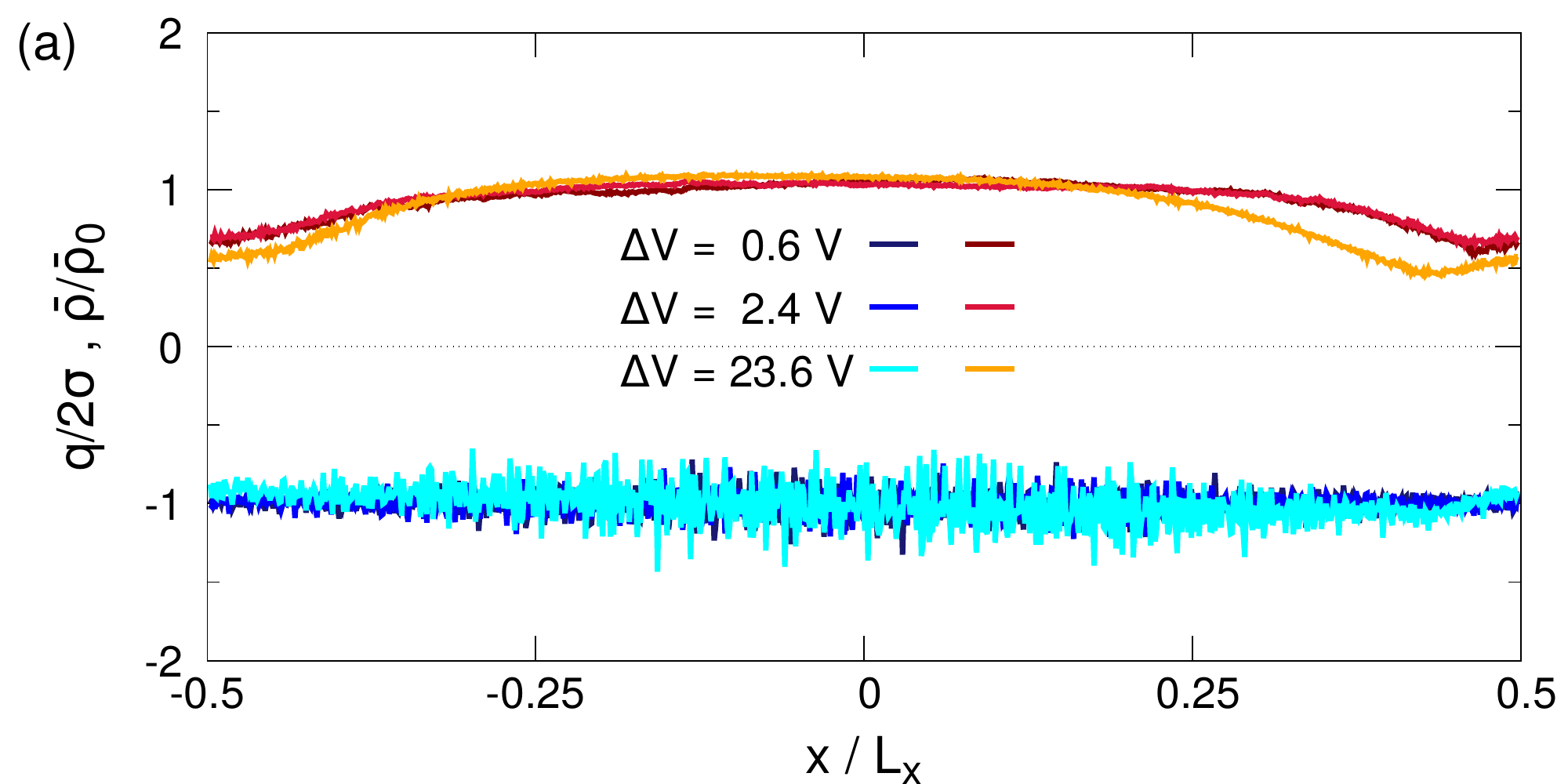}
 \includegraphics[width=0.4\textwidth]{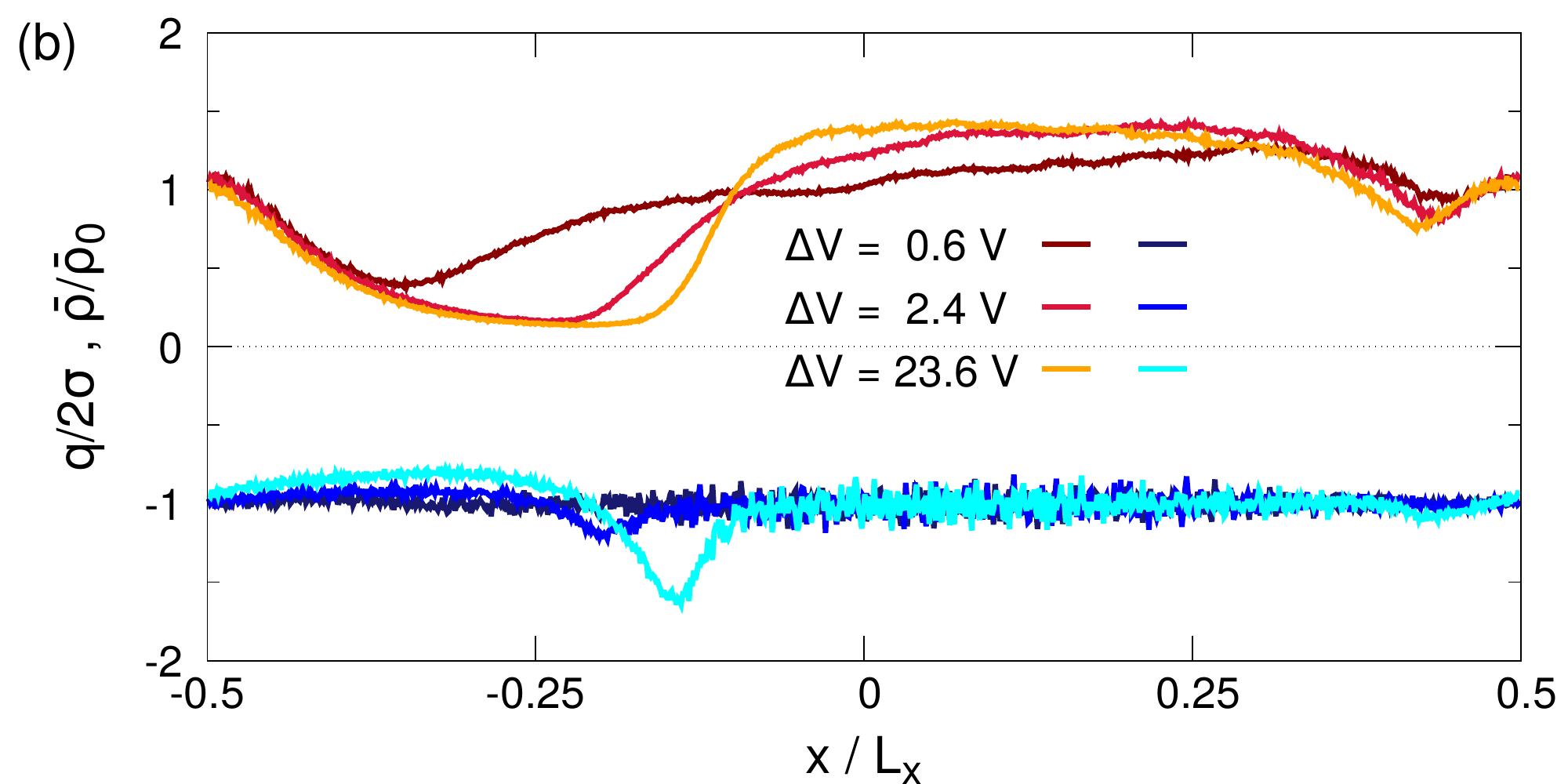}
 \includegraphics[width=0.4\textwidth]{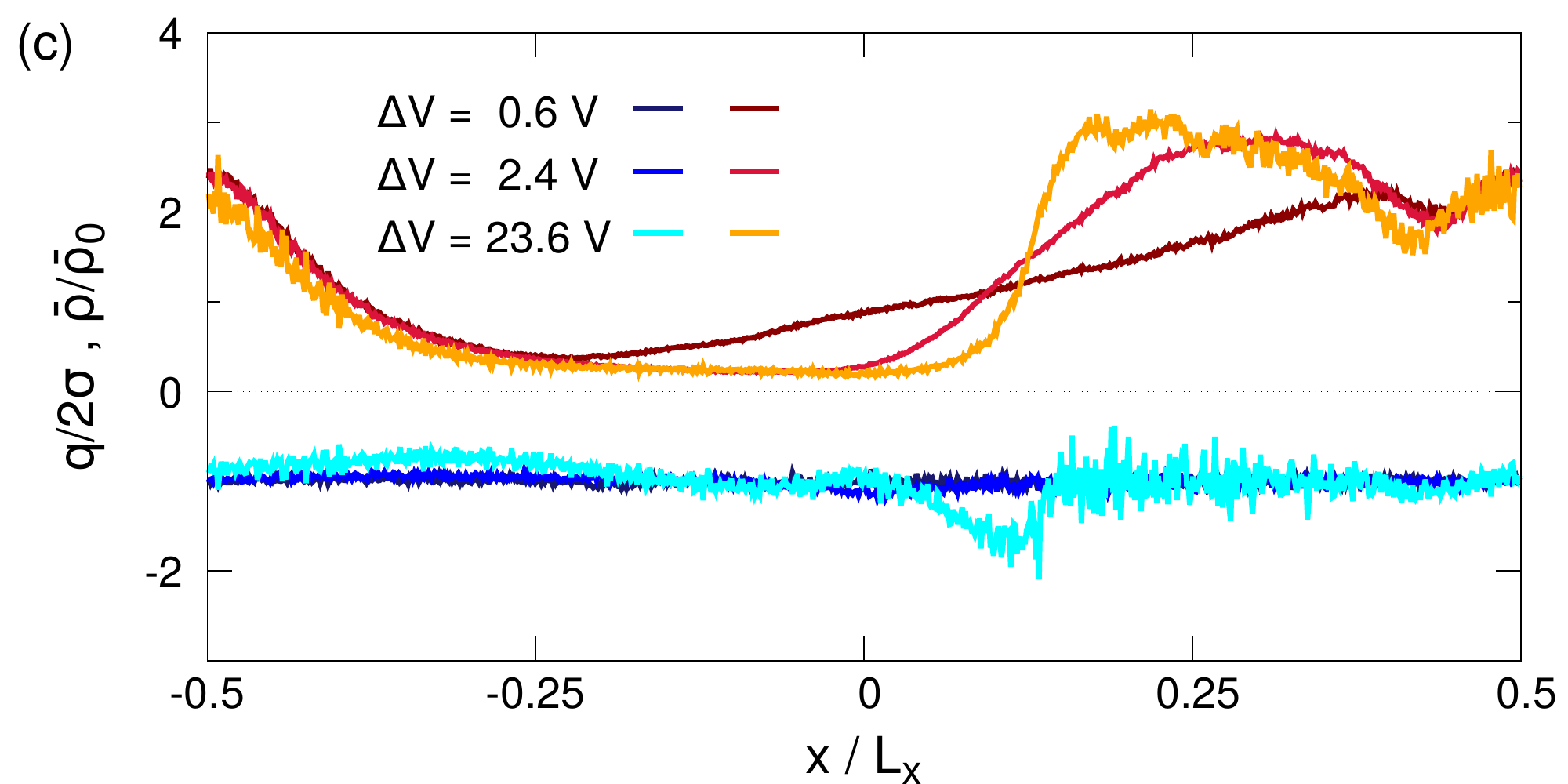}
 \caption{
 $\bar\rho(x)$ (red curves) and $q(x)$ (blue curves), defined respectively in Eq.~(\ref{eq:def-q}) and Eq.~(\ref{eq:bar-rho}), as a function of the longitudinal position for $\rho=3M$ (panel a), $\rho=1 M$ (panel b) and $\rho=0.3 M$ (panel c). Lighter color stand for large values of the potential drop $\Delta V$, see legend.} 
 \label{fig:profx}
\end{figure}

In order to quantitatively capture the accumulation of ions density, 
we have analyzed the dependence of the ionic densities averaged over the channel section as a function of the 
longitudinal position.
Fig.\ref{fig:profxs} shows 
that the dependence of normalized densities profiles $\bar \rho^\pm(x)$ defined as 
\begin{equation} 
 \bar\rho_\pm(x)\equiv\frac{1}{2h(x)L_y}\int\limits_0^{L_y}dy\int\limits_{-h(x)}^{h(x)}
\rho_\pm(x,y,z) dz
\label{eq:barrho}
\end{equation}
on the longitudinal position strongly depends on the value of the Debye length as compared to the channel section at the bottleneck. Indeed, when $\lambda/h_{min}\ll1$, i.e. for $\rho_0=3M$, panel a) of Fig.\ref{fig:profxs} shows that the density profiles are almost constant along the channel for all the values of the external force we have tested.
The only relevant difference occur close to the narrowest sections $x/L_x \simeq \pm 0.5$.
 In contrast, when $\lambda/h_{min}\sim1$, i.e. for $\rho_0=1M$ and $\rho_0=0.3M$ panels b)-c) of Fig.\ref{fig:profxs} show the onset of a region where the coions ($K^+$) are depleted and the left-right symmetry is broken. 
Once we have analyzed the density of $Cl^-$ and $K^+$ separately we move to the local total charge and local ionic densities defined as:
\begin{align}
 q(x) \equiv &2h(x)\left[\bar\rho_+(x)-\bar\rho_-(x)\right]  \label{eq:def-q}\\
 \bar\rho(x) \equiv &\bar\rho_+(x)+\bar\rho_-(x)\label{eq:bar-rho}
\end{align}
We stress that we choose a different normalization for $q(x)$ and $\bar \rho(x)$ so that in 
a plane channel $|q(x)|$ always match the total surface charge of channel 
walls $2|\sigma|$ whereas $\bar \rho(x)$ is the average density that, 
in the Debye-H\"uckel regime, determines the local value of the Debye 
length\footnote{We stress that according to Eq.(\ref{eq:bar-rho}) 
we have that $\frac{1}{L_x}\int_0^{L_x}\bar\rho(x)dx\neq\bar\rho_0$ 
where\newline 
$\bar\rho_0=\dfrac{\int\limits_0^Ldx\int\limits_0^{L_y}dy\int\limits_{-h(x)}^{h(x)}\rho_+(x,y,z)+\rho_-(x,y,z)dz}{\int\limits_0^L2h(x)L_yL_xdx}$\,\,\,\,\,\,\,\,\,
is the density averaged over the full volume.}.
Interestingly, Fig.~\ref{fig:profx}.a shows that in the limit of small Debye lengths, 
$\lambda/h_{\text{min}} < 1$, local electroneutrality is recovered, i.e. $q(x)\simeq 2\sigma$,  
and the ionic density profile is almost constant along the channel.
Only at very high voltages a small deviation is present at $x/L_x \simeq 0.4$.

In contrast, for larger values of the Debye length, 
$\lambda/h_{\text{min}} \simeq 1$, neither $q(x)$ nor $\bar\rho(x)$ are constant. 
Such inhomogeneities trigger the onset of modulations in the magnitude of the 
local electrostatic potential. In particular, an accurate inspection of 
panels b)-c) of Fig.\ref{fig:profx} reveals that weaker values of the external 
field triggers solely an inhomogeneity in the ionic density but do not 
affect the local electroneutrality. Interestingly, upon increasing the strength of 
external electric field local electroneutrality 
breaks down and on the top of the well-known concentration polarization (CP)~\cite{Mani2011,Pourcelly2014,Yossifon2014,Leese2014,Andersen2017},
 a charge polarization (QP) appears.
We remark that charge polarization sets for smaller values of the external force for smaller values of the Debye length. This can be due to finite liquid slippage at solid wall that cannot be disregarded in the regime under study.
Indeed, slippage is commonly described in term of the Navier boundary condition
that, for a plane channel, reads
$u_t \vert_{wall} = L_s \partial u_t / \partial n$   
with $u_t$ the component of the velocity field tangent to the wall, $n$ the 
normal to the wall and $L_s$ the slip length~\cite{lauga2007microfluidics}.
Atomistic simulation showed that 
for smooth hydrophilic and slightly hydrophobic 
(contact angle $<120^\circ$)
surfaces $L_s$, hardly exceeds a nanometer~\cite{chinappi2010intrinsic,huang2008water,sega2013regularization}.
In addition, the presence of a strong surface charge, further
reduces the slip length for hydrophobic surfaces~\cite{huang2008aqueous}.
\textcolor{black}{When comparing simulations performed with different ionic concentrations it should be taken into account that the slip length $L_s$ in the three setups may be slightly different. Indeed, the relevant parameter ruling the effect of slippage on the electroosmotic flow is the ratio $L_s/\lambda_D$~\cite{bocquet2010nanofluidics}. This feature is emphasized by Fig.S1 in Suppl. Mat. that shows that the mismatch between the prediction of the analytical model (see Ref.~\cite{Malgaretti2014}) and the numerical results increases upon decreasing $\lambda_D$.}
 

In conclusion, we have reported on numerical simulations concerning a $KCl$ solution embedded in a varying-section channel under the action of a constant electrostatic field. 
Our simulations show that, when the Debye length is comparable to the width of the channel 
bottlenecks, the system is in the entropic electrokinetic regime
that is characterized by the onset of eddies~\cite{Malgaretti2014}. 
In this perspective we observe, in agreement with what has been reported in the literature, 
the onset of a concentration polarization and local recirculation of the fluid velocity 
that comes along with the onset of a standing {\sl shock} 
in the ionic concentration. 
Surprisingly, for stronger external fields the local electroneutrality breaks 
down and an additional charge polarization (QP) sets in. Such a novel phenomena has been observed thanks to our microscopic approach based on Molecular Dynamics simulations in which the ionic densities are not constrained. 
In this perspective, our results show that for mild external fields local 
electroneutrality is recovered. This can justify \textit{a posteriori} 
the assumption of local electroneutrality in these regimes. 
However, for larger external fields, local 
electroneutrality does not hold and a net electric dipole sets in inside the channel.

\paragraph*{Acknowledgments}
PM acknowledges Dr. Mathijs Janssen for useful discussions. This research used the computational resource from CINECA (NATWE project), and the Swiss National Super-computing Centre (CSCS), project ID sm11. 

\appendix
\section{Ion and electroosmotic currents}

\begin{figure}[]
 \includegraphics[width=0.48\textwidth]{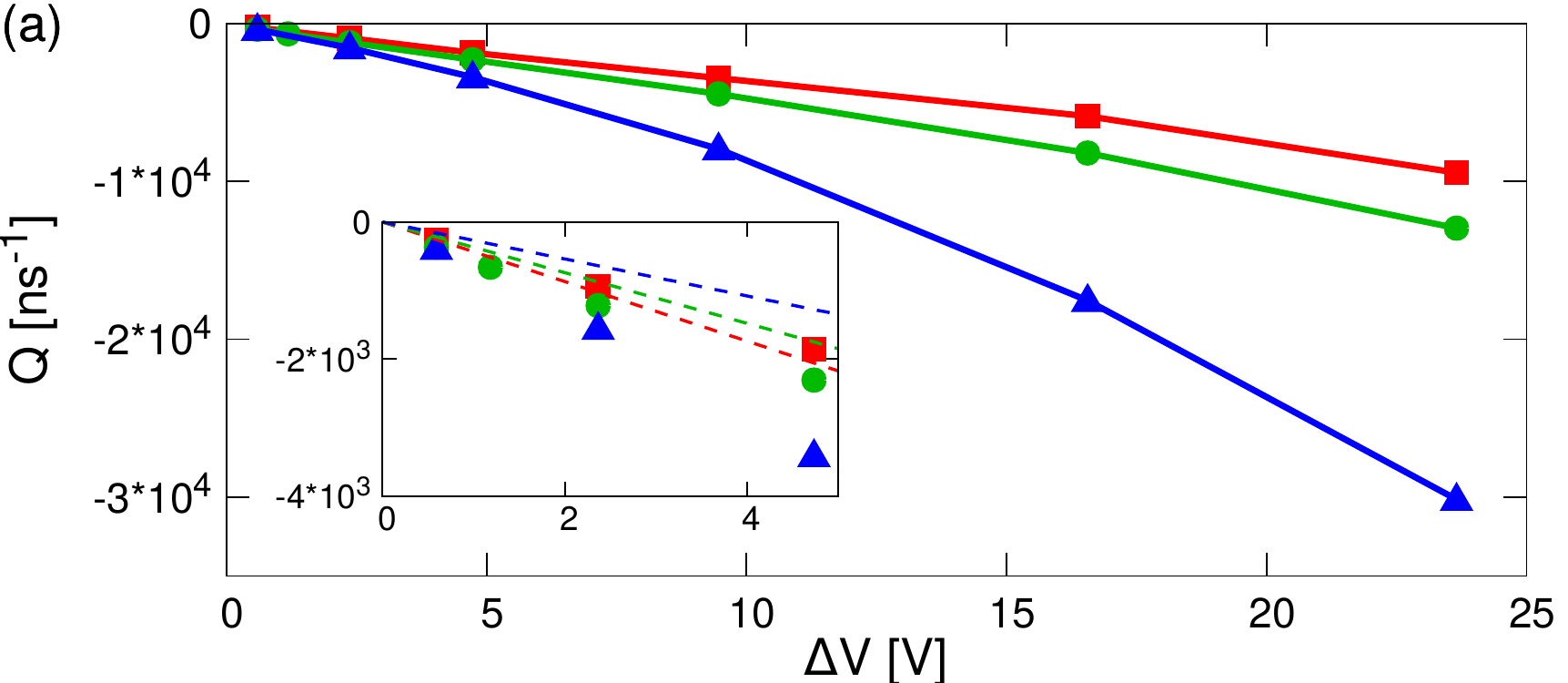}
 \includegraphics[width=0.48\textwidth]{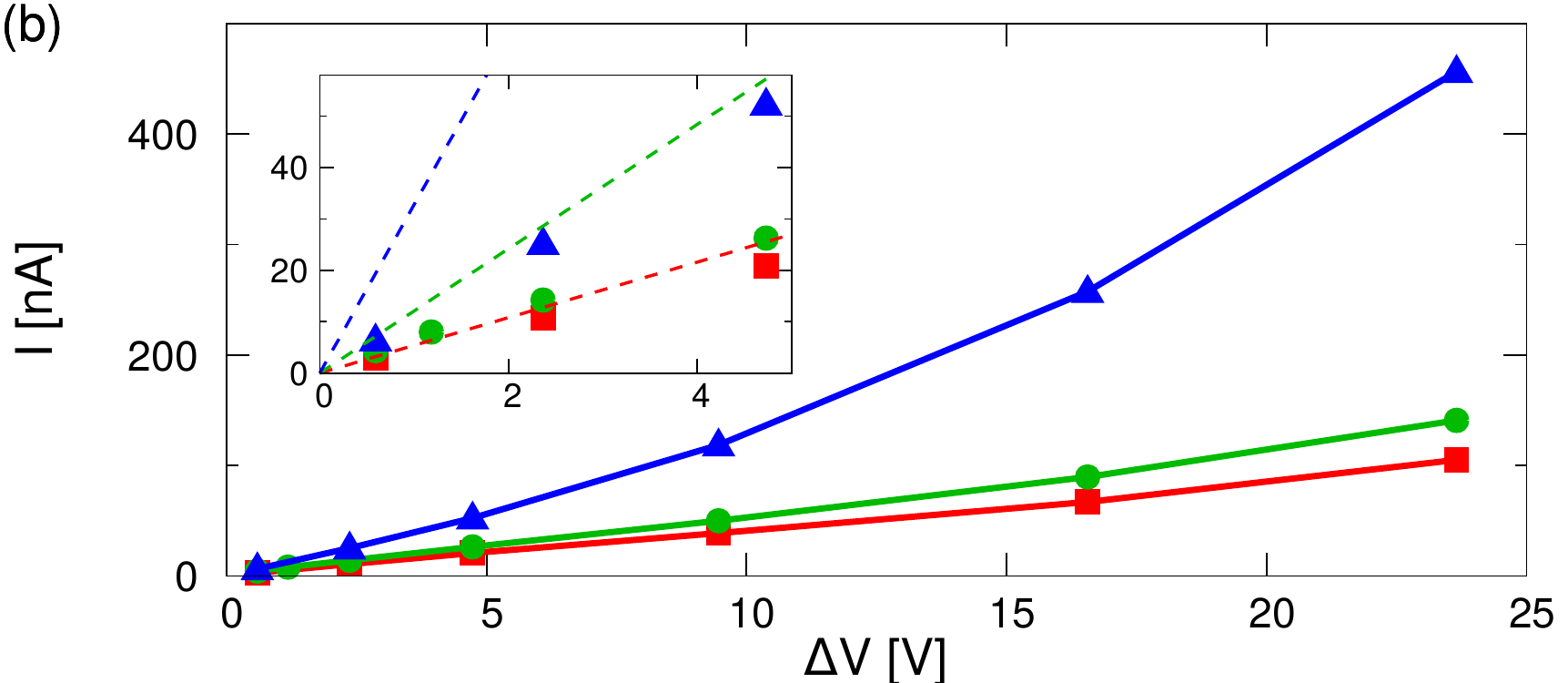}
 \caption{ \label{fig:transport}
 Electrolyte transport. a): Electro-osmotic flux $Q$ expressed in water molecules per $ns$
  as function of $\Delta V$. 
 Different colors stand for different concentrations, namely $\rho_0=0.3$M (green circles)
$\rho_0=1$M (red squares) and $\rho_0=3$M (blue triangles).
 b):  Ionic current
for different electrostatic potential drops $\Delta V$. Ionic concentrations are color-coded as in panel a).
The insets of both panels show the PNP prediction (dashed lines) in the low voltage
regime.
}
\end{figure}

Here, we discuss the dependence of ionic and electroosmotic currents on the applied voltage. 
Figures~\ref{fig:transport}.a and \ref{fig:transport}.b show the electro-osmotic flow $Q$ and the ionic current $I$ as a function of the applied electrostatic potential drop $\Delta V = L_x E_x$ along with the predictions of Poisson-Nernst-Plank (PNP) theory presented in~\cite{Malgaretti2014}. 
Both panels of Fig~\ref{fig:transport} show a significant mismatch between the numerical results (symbols) and the PNP model (dashed lines).
Interestingly the mismatch increases upon increasing the salt concentration, i.e. decreasing the Debye length. As discussed in the main text, the effect of the slip on the walls of the channel increases as the ratio $L_s/\lambda_D$ where $L_s$ is the so-called slip length and it captures the magnitude of the slip. In this perspective, Fig.\ref{fig:transport}.a makes us speculate that the mismatch between the numerical results and the PNP model is due to the partial slip of the fluid on the channel walls. Indeed, for high concentration (low $\lambda_D$), the MD electro-osmotic flux (blue triangles) is significantly larger than the PNP prediction (dashed blue line).
The partial slip can also explain the mismatch shown in Fig.\ref{fig:transport}.b. While in a flat channel an enhancement of the volumetric fluid flow $Q$ leads to an increased electric current (see Ref.~\cite{bruus2008theoretical}) the situation is different in the case of a corrugated channel. Indeed, in the latter case the $K^+$ ions have to overcome a potential barrier at the channel bottleneck whereas $Cl^-$ can move along the channel walls at almost constant potential. 
Therefore the enhanced $Q$ will reduce the magnitude of the effective barrier that $K^+$ have to overcome. Since the flow is quite sensitive to the high of the barrier (for very large barrier the time scales 
\textit{exponentially} with the high of the barrier) the enhancement of $Q$ will be more beneficial to the flow of $K^+$ as compared to the gain for $Cl^-$. Since the electric current is defined as $I=J_K-J_{Cl}$ the enhancement of $Q$ leads to a hampered electric current, i.e. the electric currents from MD (symbols) are smaller than the PNP predictions (dashed lines), in particular for larger ionic concentrations.

\section{Methods}

{\bf System set-up and equilibration.}
The system is constituted by a KCl
water solution confined by two solid walls. 
The solid walls are formed by atoms 
arranged in a face centered cubic (fcc) structure with lattice 
side $a = 4.92$ \AA.
The mass of each atom is $m_w = 70\; a.m.u.$ and they interact 
via Lennard-Jones potential with 
potential well depth
$\epsilon_w = 0.152$ Kcal/mol) 
and van der Waals radius $\sigma_w = 3.15$ \AA.
The curved walls are obtained by solid 
slabs cut parallel to 111 planes of the fcc solid.
The two slabs are then bend to follow the 
prescribed channel shape with a proper change of the z-coordinate.
Each atom exposed to the liquid
has a charge $q_w$. 

Water molecules and ions at a molar concentration $\rho_0$ are added using 
VMD~\cite{humphrey1996vmd}.
In particular, indicated with  
$N_{Cl^-}$ and $N_{K^+}$ the number of Chloride and Potassium ions, we 
have
$N_{Cl^-}=1749$ and $N_{K^+}=789$ 
for $\rho_0 = 0.3 M$,
$N_{Cl^-}=3327$ and $N_{K^+}=2367$ 
for $\rho_0 = 1 M$,
and 
$N_{Cl^-}=8062$ and $N_{K^+}=7102$ 
for $\rho_0 = 3 M$.
In each case, $N_{Cl^-} - N_{K^+} = 960 $.
Since $960$e is the total charge on the walls, the total charge of the system is zero.

All MD simulations are performed using NAMD 
software~\cite{phillips2005scalable} by implementing periodic boundary conditions 
being $L_x=836.6$~\AA, $L_y=60.37$~\AA~ and $L_z=200$~\AA~ the box sizes.
$L_x$ and $L_y$ are dictated by the size of solid walls while
$L_z$ is arbitrary set to $200$~\AA~ to avoid 
non-bonded interactions among the different images along the z-axes.
The total number of atoms is $\sim$400 000. 
A snapshot of the system is reported in figure~1.b.
Particle mesh Ewalds (PME) summation method was employed 
for the electrostatics~\cite{batcho2001optimized}.  
TIP3P model~\cite{jorgensen1983comparison}
 was employed for water while
CHARMM36 force field~\cite{brooks2009charmm,vanommeslaeghe2012automation} 
with NBFIX corrections 
is used for the ions~\cite{luo2009simulation}.
More accurate water models are available, 
see e.g. the TIP4P\_2005~\cite{abascal2005general}
often used for nanofluidics~\cite{gentili2014pressure,tinti2017intrusion}.
However, these models are much more computational demanding than TIP3P,
hence, in this work, beside the well known limitations of TIP3P 
(e.g. the viscosity is lower than the experimental value)
 we preferred TIP3P since it allowed us to run longer simulations
and to partially reduce the statistical error.

Walls atoms are constrained at their initial position by 
harmonic springs (spring constant $k=10$ Kcal/(mol \AA$^2$)). 
A first $0.8$ ns NVT simulations ($T=310$K, time step  $2$ fs)
was run and pressure is estimated from the average force acting on the 
walls. The distance between the upper and lower wall 
is then gradually reduced 
and further $0.8$ ns simulations are run until 
the pressure reached a value close to $1$ atm.

{\bf Production runs.}
As common in the MD studies on ionic and electroosmotic transport, 
a homogeneous and constant 
external electric field ${\bf E} = (E_x,0,0)$ is applied 
to the whole system~\cite{aksimentiev2005imaging,wilson2016graphene,bonome2017electroosmotic}.
Snapshots are acquired every $\Delta t = 2$ps. 

The water velocity field in the x,z plane is calculated as follows.
The velocity ${\bf v_{i}}$ of the i-th water molecule
is estimated as
${\bf v}_{i} =  [{\bf x}_{i} (t + \Delta t) - {\bf x_{i}}(t) ]/\Delta t$
where ${\bf x_{i}}$ is the position of the oxygen atom of the i-th water
molecule. 
The mean velocity of each particle ${\bf v_{i}}$
is associated to the point 
${\bf \hat  x}_{i} = [ {\bf x}_{i} (t + \Delta t) + {\bf x}_{i}(t)]/2$.
The value of the velocity field ${\bf u}$ in a given point ${\bf x}$ is 
hence calculated as the mean of the velocities 
of the particles in the neighborhood of $\bf x$, i.e. the molecule 
for which ${\bf \hat x_{i}}$
belongs to a region $[x - \Delta x/2, x + \Delta x/2] 
\cap [z - \Delta z/2, z + \Delta z/2]$.   
The production runs
span in a range of $40-90$ ns with longer simulations used for smaller forcing and lower
ionic concentration.
\bibliography{biblio}
\end{document}